\documentclass[manuscript,acmsmall,screen]{acmart}

\usepackage{textcomp}
\usepackage{xcolor}
\usepackage{import}
\usepackage{hyperref}
\usepackage{tabularx}
\usepackage{multirow}
\usepackage{algorithm}
\usepackage[noend]{algpseudocode}
\usepackage{amsmath}
\usepackage{graphicx}
\usepackage{subcaption}
\usepackage{mathtools}
\usepackage{booktabs} 

\usepackage{supertabular}
\usepackage{float}
\usepackage{float}

\usepackage{tikz}

\acmJournal{TECS}
\acmYear{2024} \acmVolume{1} \acmNumber{1} \acmArticle{1} \acmMonth{1}

\begin{document}

\title{Topology-aware Detection and Localization of Distributed Denial-of-Service Attacks in Network-on-Chips}

% \thanks{This work was partially supported by NSF grant SaTC-1936040}
% }

\author{Hansika Weerasena}
\affiliation{%
  \institution{University of Florida}
  \city{Gainesville}
  \state{FL}
  \postcode{32611}
  \country{USA}
}
\email{hansikam.lokukat@ufl.edu}

\author{Xiaoguo Jia}
\affiliation{%
  \institution{University of Florida}
  \city{Gainesville}
  \state{FL}
  \postcode{32611}
  \country{USA}
}
\email{xiaoguojia@ufl.edu}

\author{Prabhat Mishra}
\affiliation{%
  \institution{University of Florida}
  \city{Gainesville}
  \state{FL}
  \postcode{32611}
  \country{USA}
}
\email{prabhat@ufl.edu}

\begin{abstract}

Network-on-Chip (NoC) enables on-chip communication between diverse cores in modern System-on-Chip (SoC) designs. With its shared communication fabric, NoC has become a focal point for various security threats, especially in heterogeneous and high-performance computing platforms. Among these attacks, Distributed Denial of Service (DDoS) attacks occur when multiple malicious entities collaborate to overwhelm and disrupt access to critical system components, potentially causing severe performance degradation or complete disruption of services. These attacks are particularly challenging to detect due to their distributed nature and dynamic traffic patterns in NoC, which often evade static detection rules or simple profiling. This paper presents a framework to conduct topology-aware detection and localization of DDoS attacks using Graph Neural Networks (GNNs) by analyzing NoC traffic patterns. Specifically, by modeling the NoC as a graph, our method utilizes spatiotemporal traffic features to effectively identify and localize DDoS attacks. Unlike prior works that rely on handcrafted features or threshold-based detection, our GNN-based approach operates directly on raw inter-flit delay data, learning complex traffic dependencies without manual intervention. Experimental results demonstrate that our approach can detect and localize DDoS attacks with high accuracy (up to 99\%) while maintaining consistent performance under diverse attack strategies. Furthermore, the proposed method exhibits strong robustness across varying numbers and placements of malicious IPs, different packet injection rates, application workloads, and architectural configurations, including both 2D mesh and 3D TSV-based NoCs. Our work provides a scalable, flexible, and architecture-agnostic defense mechanism, significantly improving the availability and trustworthiness of on-chip communication in future SoC designs.

\end{abstract}




\maketitle

% \title{\huge Topology-aware Detection and Localization of Distributed Denial-of-Service Attacks in NoCs using Graph Neural Networks}

% \title{\huge Detection and Localization of Distributed Denial-of-Service Attacks in NoCs using Graph Neural Networks}

% \title{\huge Detection and Localization of Denial-of-Service Attacks in Network-on-Chips using Graph Neural Networks}

%\thanks{This work was partially supported by NSF grant SaTC-1936040}

% \author{\IEEEauthorblockN{Hansika Weerasena and Prabhat Mishra}
% \IEEEauthorblockA{Department of Computer \& Information Science \& Engineering\\
% University of Florida, Gainesville, Florida, USA}}

% \author{Hansika~Weerasena,~\IEEEmembership{Student Member,~IEEE,}
% and~Prabhat~Mishra,~\IEEEmembership{Fellow,~IEEE,}% <-this % stops a space
% \IEEEcompsocitemizethanks{\IEEEcompsocthanksitem H. Weerasena, and P. Mishra are with the Department of Computer \& Information Science \& Engineering, University of Florida, Gainesville, Florida, USA.}
% }

% \begin{IEEEkeywords}
% network-on-chip, distributed-denial-of-service, on-chip communication security, graph neural networks.
% \end{IEEEkeywords}

% \pagestyle{empty}

\section{Introduction}
\label{sec:introduction}

\begin{figure}[tp]
\centering
% \vspace{-0.2in}
\includegraphics[width=0.8\columnwidth]{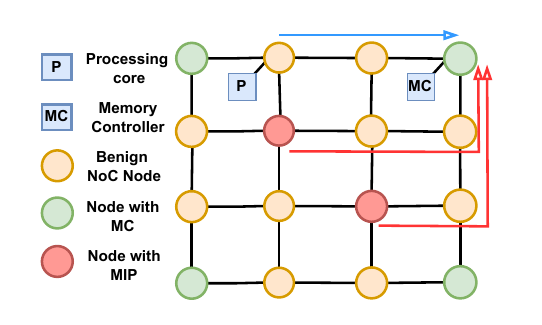}
\vspace{-0.2in}
\caption{4x4 mesh NoC topology. Only corner nodes have shared Memory Controllers (MCs). Malicious IPs (MIPs) can target MCs and flood packets to launch DDoS attack.}
\label{fig:intro}
% \vspace{-0.2in}
\end{figure}

Parallel workloads and specialized computing accelerators, such as neural network accelerators, have become pivotal in advancing computing capabilities, significantly shaping modern processor architecture designs. These workloads demand fast, energy-efficient communication across a growing number of cores, often under tight latency constraints. Heterogeneous Systems-on-Chips (SoCs) and Multi-Processor SoCs (MPSoCs) now integrate a large number of Intellectual Property (IP) cores in a single chip, reflecting a major shift towards more complex and powerful systems with diverse compute and memory hierarchies. For example, Intel's Xeon® Scalable Processor~\cite{Intelxeon} supports up to 128 cores, and Altra® multicore server processors feature up to 192 cores~\cite{amepere}. Recent advancements in heterogeneous integration and 3D chip stacking are expected to push this boundary even further~\cite{lundstrom2022moore}, enabling ultra-dense core integration on future SoCs. To support this growing communication demand, Network-on-Chip (NoC) has emerged as the de facto standard for inter-core communication, meeting the performance, scalability, and modularity requirements of densely packed chips. NoC ensures high throughput and low latency by offering a robust, scalable, and distributed communication framework within the chip. For instance, Intel employs the Skylake Mesh NoC~\cite{Intelxeon} in their server-grade processors to efficiently manage communication across many cores. Figure~\ref{fig:intro} illustrates a $4\times4$ mesh NoC comprising 16 IP cores. Each node in the NoC contains a router, a Network Interface (NI), and an IP core. When a core needs to communicate with another, it injects packets through its NI, which are routed hop-by-hop through intermediate routers to reach the destination. For example, when a processing core (\textit{P}) initiates a memory request after a cache miss, it forwards the request to one of the four corner memory controllers, as shown in Figure~\ref{fig:intro}, highlighting the fundamental role of NoC in memory access and inter-core communication.

Due to cost and time-to-market constraints, SoC manufacturers often use third-party vendors and services from the global supply chain~\cite{weerasena2024security}. Typically, only a few IP cores are designed in-house, while the others are reusable IPs from third-party vendors. This long supply chain introduces the risk of malicious implants through various channels, such as untrustworthy CAD tools, rogue designers, or foundries. Furthermore, the growing complexity of SoC designs makes comprehensive security verification increasingly difficult. Microelectronics and Advanced Packaging Technologies (MAPT) roadmap~\cite{SRC_MPAT_Roadmap} by Semiconductor Research Corporation (SRC) introduces Heterogeneous Integration as the key concept for cost- and power-efficient design in next-generation computing systems. Heterogeneous Integration increases security concerns by expanding the attack surface due to its complexity. The scalability of Heterogeneous Integration enables the integration of on-chip Machine Learning (ML) accelerators, which can be utilized for ML-based traffic monitoring to enhance security. While designing energy-efficient NoCs is a primary goal, securing them is equally important since exploiting an NoC could give attackers access to shared inter-core communications and the ability to compromise the security of the entire computing infrastructure.

% For example, FlexNoc interconnect is used by four out of the top five fabless companies to facilitate their on-chip communication \cite{js2015runtime}.

% Security threats across different NoC technologies (electrical, wireless, optical, and hybrid) have been widely explored in the literature~\cite{weerasena2024security}.Denial-of-Service (DoS) attacks target critical resources, overwhelming communication and processing, ultimately disrupting access to system services. Distributed DoS (DDoS) attacks leverage multiple malicious nodes, making them more effective and stealthy. Figure~\ref{fig:into} illustrates a DDoS attack where two malicious IPs (MIPs) overwhelm a shared Memory Controller (MC) by flooding it with bogus memory requests. Previous work~\cite{js2015runtime,charles2020real} on mitigating DoS attacks in NoCs have focused on statistical traffic modeling, but these methods rely on statistical bounds, which fail to adapt to changing application characteristics. While existing traditional machine learning (ML)-based mitigation methods are promising~\cite{sudusinghe2021denial,sinha2021sniffer}, they require extensive manual feature engineering and retraining as applications evolve. Moreover, some of these approaches~\cite{sudusinghe2021denial,js2015runtime} are limited to detecting DoS attacks with a single adversary. Recent advances in acceleration of deep neural networks (DNNs) and heterogeneous integration with chiplets enable on-chip DNN acceleration for real-time traffic monitoring. 

% \subsection{State-of-the-Art and its Limitations}

Security threats across different NoC technologies (electrical, wireless, optical, and hybrid) have been widely explored in the literature~\cite{weerasena2024security}. Denial-of-Service (DoS) attacks target critical resources, overwhelming communication and processing, ultimately disrupting access to system services. Distributed DoS (DDoS) attacks leverage multiple malicious nodes, making them more effective and stealthy. Such attacks are especially problematic in systems with shared memory controllers, caches, or accelerators, where a bottleneck at a single point can cascade into global slowdowns. Figure~\ref{fig:intro} illustrates a DDoS attack where two malicious IPs (MIPs) overwhelm a shared Memory Controller (MC) by flooding it with bogus memory requests. Previous work~\cite{js2015runtime,charles2020real} on mitigating DoS attacks in NoCs have focused on statistical traffic modeling, but these methods rely on statistical bounds, which fail to adapt to changing application characteristics. While existing traditional machine learning (ML)-based mitigation methods are promising~\cite{sudusinghe2021denial,sinha2021sniffer}, they require extensive manual feature engineering and retraining as applications evolve. Furthermore, their reliance on handcrafted thresholds or congestion metrics makes them less effective under dynamic, stealthy DDoS behaviors. Moreover, some of these approaches~\cite{sudusinghe2021denial,js2015runtime} are limited to detecting DoS attacks with a single adversary. They also do not leverage the topological structure of NoC, which encodes spatial dependencies vital for effective threat modeling. Recent advances in acceleration of deep neural networks (DNNs) and heterogeneous integration with chiplets enable on-chip DNN acceleration for real-time traffic monitoring. This presents an opportunity to combine hardware-accelerated inference with deep learning models for adaptive and scalable runtime security enforcement in future NoC-based SoCs.

In this paper, we utilize graph neural networks (GNNs) for scalable and topology-aware detection and localization of DDoS attacks in Network-on-Chip systems. Specifically, this paper makes the following major contributions:

\begin{itemize}
    \item We propose a novel methodology to detect and localize DDoS attacks in NoC architectures using GNNs, leveraging raw inter-flit delays and spatial graph structure to extract spatiotemporal features without manual tuning.

    \item The proposed GNN-based detection and localization framework is rigorously evaluated across a wide range of traffic configurations, 2D and 3D architectures, attack scenarios, and benchmark applications, demonstrating its generalization and scalability
\end{itemize}

This paper is organized as follows. Section~\ref{sec:related_work} provides background on core concepts and surveys related efforts. Section~\ref{sec:threat} outlines the threat model. Section~\ref{sec:methodology} describes our proposed framework for detection and localization of DDoS attacks. Section~\ref{sec:experiments} presents the experimental results and discusses their implications. Finally, Section~\ref{sec:conclusion} concludes the paper.

\section{Background and Related Work}
% \label{sec:background}
\label{sec:related_work}

This section provides background on Network-on-Chip architectures, multivariate time series, and Graph Neural Networks. It also reviews existing techniques for detecting and mitigating security threats, with a focus on denial-of-service and distributed denial-of-service attacks in NoCs, highlighting their limitations and motivating the need for a topology-aware solution.

\subsection{Network-on-Chip and Communication Protocol}

\begin{figure}[tp]
\centering
% \vspace{-0.2in}
\includegraphics[width=0.7\columnwidth]{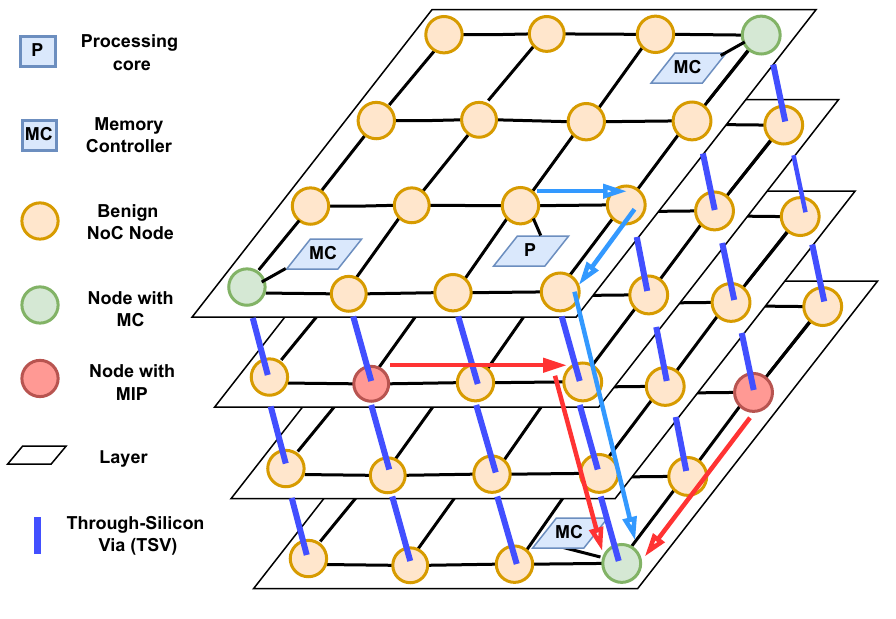}
% \vspace{-0.2in}
\caption{4x4x4 3D mesh NoC topology with TSVs for vertical communication.}
\label{fig:3D_NoC}
% \vspace{-0.2in}
\end{figure}

Network-on-Chip (NoC) has become the de facto communication backbone for modern Multi-Processor Systems-on-Chip (MPSoCs) due to its scalability, modularity, and performance advantages. The most common topology is a 2D mesh NoC, which consists of a grid of nodes—each node typically comprising a router, a network interface (NI), and a processing element or IP core. The NI is responsible for converting high-level messages from the core into packets, which are further broken down into smaller units called flits (flow control digits). These flits are injected into the NoC and traverse hop-by-hop through routers toward their destination. Routers are lightweight switch-like components that forward flits based on routing algorithms. The most common routing mechanism in NoC is deterministic XY routing, which first routes in the X-direction and then in the Y-direction. This routing scheme is simple to implement and ensures deadlock avoidance.

Beyond 2D meshes, several 3D NoC topologies have emerged to enhance performance, scalability, and bandwidth utilization. In a 3D NoCs, multiple 2D layers are stacked vertically and connected via Through-Silicon Vias (TSVs)~\cite{motoyoshi2009through}, enabling high-bandwidth, low-latency communication across layers. Figure~\ref{fig:3D_NoC}shows a typical 3D NoC with TSVs for vertical connections. TSVs act as vertical buses that link corresponding routers in adjacent layers, significantly reducing inter-layer hop counts and wire length. To manage access contention across layers, 3D NoCs often implement bus arbitration mechanisms to select a sender when multiple layers attempt to transmit over the same TSV simultaneously~\cite{tatas2014designing}. Routing in 3D NoCs extends conventional XY routing by introducing Z-direction hops either before or after XY traversal, depending on architectural constraints and optimization goals. These vertically integrated designs introduce new opportunities for performance and energy efficiency but also present increased complexity and new security vulnerabilities due to denser connectivity.  Compared to traditional off-chip networks, NoCs are designed to be lightweight and fast, with minimal protocol overhead. Unlike general-purpose networks, NoCs operate under tight area and power constraints, and their communication patterns are usually more predictable. As a result, NoCs can achieve high throughput and low latency with simpler protocols. However, their centralized nature and shared interconnects also make them vulnerable to a range of security threats, including denial-of-service attacks.

\subsection{Multivariate Time Series}

Time series data represent sequences of observations collected over time. In a \textit{univariate time series}, each time step contains a single scalar value, i.e., \( \mathbf{X} = \{x_1, x_2, \ldots, x_T\} \in \mathbb{R}^T \), where each \( x_t \in \mathbb{R} \). These values may be sampled at uniform or non-uniform time intervals. In contrast, a \textit{multivariate time series} consists of multiple variables (or dimensions) observed simultaneously at each time step. This results in a matrix \( \mathbf{X} \in \mathbb{R}^{N \times T} \), where each column \( \mathbf{x}_t \in \mathbb{R}^N \) represents the observations of all \( N \) variables at time \( t \). In the context of NoCs, multivariate time series can represent traffic metrics such as packet delay, flit injection rate, or buffer occupancy which are collected over time at each node or router in the network. This format allows temporal trends (e.g., rising congestion) and spatial correlations (e.g., traffic ripple effects across nodes) to be jointly analyzed. Multivariate time series are commonly used in graph-based models due to their ability to capture both temporal dynamics (e.g., changes over time within each variable) and spatial correlations (e.g., relationships between variables or nodes at a single time step). These temporal and spatial dependencies can be effectively modeled using spatial-temporal graphs, making them ideal inputs for Graph Neural Networks, as discussed next.

\subsection{Graph Neural Networks}
\label{sec:gnn_background}

Graphs are a powerful abstraction to represent relational structures among entities. A graph is defined as \( \mathcal{G} = (\mathcal{V}, \mathcal{E}) \), where \( \mathcal{V} = \{v_1, v_2, \ldots, v_N\} \) is the set of \( N \) nodes, and \( \mathcal{E} \subseteq \mathcal{V} \times \mathcal{V} \) is the set of edges. An \textit{attributed graph} augments this structure by associating each node \( v_i \) with a feature vector \( \mathbf{x}_i \in \mathbb{R}^D \). The graph is then represented as \( \mathcal{G} = (\mathbf{A}, \mathbf{X}) \), where \( \mathbf{A} \in \mathbb{R}^{N \times N} \) is the (possibly weighted) adjacency matrix encoding the connectivity of the graph, and \( \mathbf{X} \in \mathbb{R}^{N \times D} \) is the node feature matrix. 

In dynamic settings, where node features evolve over time, a sequence of attributed graphs \( \mathcal{G}_t = (\mathbf{A}_t, \mathbf{X}_t) \) for \( t = 1, \ldots, T \) can be used to represent spatiotemporal graphs. These structures capture both inter-variable dependencies (through the edges of the graph) and temporal dynamics (through evolving node features over time). The adjacency matrices \( \mathbf{A}_t \) may remain fixed or vary with time, depending on the specific application. Graph Neural Networks (GNNs) are a class of deep learning models that operate on such graph-structured data. A GNN learns node embeddings by iteratively aggregating information from each node's neighbors. At the \( k \)-th layer of a GNN, the embedding of node \( v_i \) is updated through two primary functions: \textsc{Aggregate} and \textsc{Combine}. The update can be expressed as:

\[
\mathbf{a}_i^{(k)} = \textsc{Aggregate}^{(k)} \left( \left\{ \mathbf{h}_j^{(k-1)} : v_j \in \mathcal{N}(v_i) \right\} \right),
\quad
\mathbf{h}_i^{(k)} = \textsc{Combine}^{(k)} \left( \mathbf{h}_i^{(k-1)}, \mathbf{a}_i^{(k)} \right),
\]

Here, \( \mathbf{h}_i^{(k)} \) is the embedding of node \( v_i \) at layer \( k \), and \( \mathcal{N}(v_i) \) denotes the set of neighbors of node \( v_i \). The \textsc{Aggregate} function collects messages from neighboring nodes, and the \textsc{Combine} function integrates this information with the node’s current state. The initial node representation is \( \mathbf{h}_i^{(0)} = \mathbf{x}_i \), and the final output after \( K \) layers is \( \mathbf{h}_i^{(K)} \). This formulation corresponds to \textit{spatial GNNs}, where convolution operations are defined in the node domain using message passing. An alternative class, \textit{spectral GNNs}, defines convolution in the frequency domain using spectral graph theory. However, spatial GNNs are more flexible and scalable for practical applications.

When applied to time-series data, GNNs require a graph structure that encodes the dependencies between variables. If no explicit graph is available, the structure can be inferred from data using heuristics or learning-based methods. Once the graph is defined, spatial-temporal GNNs can jointly model both temporal evolution and spatial correlations, enabling powerful and expressive representations for complex systems. GNNs are well-suited for applications that require relational reasoning or topology-aware learning. This includes social networks, molecular graphs, traffic networks, and more recently, NoCs. In the context of NoC security, modeling the chip as a graph where routers are nodes and links are edges allows GNNs to naturally capture both spatial (topological) and temporal dynamics of on-chip communication.

\color{black}

\subsection{Related Work}

NoC security attacks and defenses can be categorized based on the targeted security requirements by the attacker: confidentiality, integrity, anonymity, authenticity, availability, and freshness~\cite{weerasena2024security}. Examples of attacks include attack on anonymity~\cite{sarihi2021securing, weerasena2024breaking, ahmed2020defense, charles2020lightweight}, snooping attacks~\cite{ancajas2014fort,sepulveda2017towards,raparti2019lightweight, js2015runtime, weerasena2021lightweight}, side-channel attacks~\cite{wang2012efficient, sepulveda2014noc, sepulveda2016dynamic, reinbrecht2016gossip, boraten2018securing}, and spoofing attacks~\cite{sepulveda2017towards, weerasena2024lightweight}.  While both DoS and DDoS attacks in NoCs have been studied~\cite{js2015runtime, sudusinghe2021denial, sinha2021sniffer, charles2020real}, existing mitigation approaches often yield suboptimal results due to inherent limitations. For example, previous defenses that rely on static profiling techniques~\cite{js2015runtime,charles2020real} require substantial human intervention and are difficult to adapt across system configurations, especially as the complexity and heterogeneity of SoC designs increase. They often depend on manually tuned statistical thresholds, which can be inaccurate in dynamic settings. For instance, Charles et al.~\cite{charles2020real} proposed statistically profiling packet arrival times at routers using an upper bound to detect and localize attacks. However, this simple upper bound is not effective for changing traffic patterns in NoC-based SoCs. It also lacks temporal flexibility, as threshold values may become obsolete under varying traffic loads  Moreover, any modification to the NoC architecture—such as changes in topology, routing policy, or traffic load—necessitates re-conducting experiments on the new NoC configuration and manually determining a new threshold, which is both time-consuming and error-prone, further limiting the practicality of such static approaches in modern NoCs.

ML-based defenses have been explored for mitigating both DoS and DDoS attacks. Sudusinghe et al.~\cite{sudusinghe2021denial} proposed a traditional ML-based approach using 17 manually engineered features to detect DoS attacks, limiting its scalability to more complex DDoS threats. Additionally, manual feature selection fails to capture the full complexity of NoC traffic, negatively affecting detection performance. Sinha et al.~\cite{sinha2021sniffer} proposed an approach using in-node perceptron models for local congestion detection and message passing to localize DoS/DDoS attacks in NoC. This approach has several limitations: (1) it uses only three engineered features with a traditional ML model, potentially missing complex traffic anomalies across the network; (2) it overlooks temporal and spatial dependencies in NoC traffic, leading to suboptimal performance in congestion detection, which is critical for effective attack localization; and (3) it relies on fixed thresholds (through weighted voting) to trigger actions based on congestion patterns, limiting adaptability to dynamic network conditions. Moreover, reliance on localized congestion detection may be insufficient in coordinated DDoS attacks where global traffic context is essential to accurately identify anomalies.

Recently, DL2Fence~\cite{wang2024dl2fence} introduced a CNN-based framework that integrates deep learning with frame fusion techniques to detect and localize denial-of-service (DoS) attacks in NoCs. While DL2Fence achieves promising results through classification and segmentation of fused spatiotemporal features, it requires structured pre-processing, assumes specific frame representations, and depends on handcrafted routing pattern generalizations. In contrast, our approach makes significantly fewer assumptions: it operates directly on raw inter-flit delay data, requires no manual feature engineering, and does not assume a particular format or granularity of input traffic. By explicitly modeling the NoC as a graph, our GNN-based framework inherently captures spatial relationships between nodes, enabling topology-aware detection and localization. Moreover, we formulate detection and localization as unified graph and node classification problems, respectively, allowing greater flexibility, robustness to attacker placement, and easier adaptation to diverse NoC configurations. This structural awareness, combined with minimal assumptions and raw feature input, enhances generalizability and positions our method for real-time security enforcement in complex SoC environments. 

To overcome these limitations, we propose a topology-aware framework that leverages GNNs to detect and localize DDoS attacks. Our approach eliminates the need for handcrafted features and instead directly learns spatial and temporal patterns from raw inter-flit delay measurements. By modeling the NoC as a graph, our GNN-based framework inherently captures spatial relationships between nodes, enabling topology-aware detection and localization. To the best of our knowledge, our work is the first to employ graph neural networks for detecting and localizing security threats in NoC-based SoCs, offering a scalable and architecture-agnostic defense.

\section{Threat Model}
\label{sec:threat}

\begin{figure}[tp]
\centering
% \vspace{-0.3in}
\includegraphics[width=0.9\columnwidth]{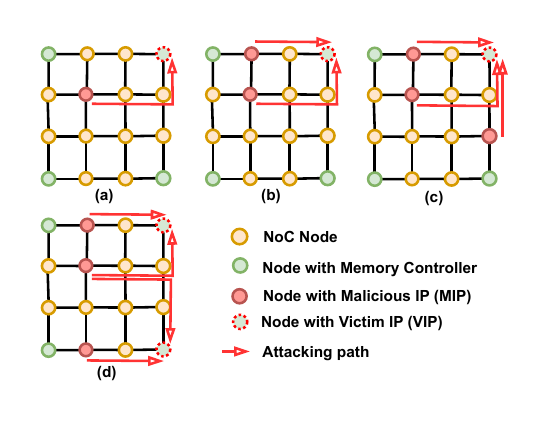}
\vspace{-0.4in}
\caption{MIP placement scenarios: (a) one MIP, (b) two MIPs attacking the same VIP, (c) three MIPs with overlapping paths targeting one VIP, (d) three MIPs attacking two VIPs.}
\label{fig:threat_model}
% \vspace{-0.2in}
\end{figure}

In NoC-based SoCs, a cache miss in a core triggers a control packet (e.g., load or store) destined for the memory controller to fetch data from main memory. In NoC communication, packets are divided into flits, which travel hop-by-hop through the network until they reach their destination (e.g., memory controller). Typically, a few memory controllers are shared across the SoC. Our threat model assumes a widely used $n \times n$ 2D mesh and $n \times n \times n$ 3D TSV-based NoC topology with $N$ total nodes. These topologies represent common architectural patterns adopted in modern many-core SoCs due to their balance between area footprint, power, and communication latency. There are four memory controllers located at the corner nodes in 2D topologies and two memory controllers for the top and bottom layer at alternating corners in 3D TSV-based topologies. In other words, there are four memory controllers shared across $N$ nodes for both cases. These shared memory resources are  attractive targets for denial-of-service attacks because they can cause system wide availability concerns.

Figure~\ref{fig:intro} shows an example scenario of a DDoS attack in a 4x4 mesh NoC where Malicious IPs (MIPs) flood bogus memory requests targeting one or more memory controllers, defined as victim IPs (VIPs). This creates traffic hotspots near critical nodes~\cite{sudusinghe2021denial}, ultimately disrupting communication within the NoC, which can result in real-time deadline violations and catastrophic failures in safety-critical systems. Such effects are particularly severe in real-time cyber-physical systems where timing constraints are critical. Our framework makes no assumptions about MIP placement, nor the number of MIPs ($N_M$) or VIPs ($N_V$). Figure~\ref{fig:threat_model} illustrates various attack scenarios involving different numbers of MIPs and VIPs, as well as different MIP placements. Figure~\ref{fig:threat_model}(a) shows a DoS attack by one MIP on one VIP. Figure~\ref{fig:threat_model}(b) and (c) represent multiple MIPs launching DDoS attacks, with (c) showing overlapping paths. Furthermore, Figure~\ref{fig:threat_model}(d) shows a scenario with two VIPs. Figure~\ref{fig:3D_NoC} illustrates a DDoS attack in a $4\times4\times4$ 3D Network-on-Chip, where multiple malicious IP cores located across two layers overwhelm the memory controller in a third layer by flooding it with excessive packets. These cases represent a wide variety of spatial placements and traffic routing interactions, making detection and localization more challenging in practice. Our proposed approach is capable of detecting and localizing DDoS attacks in all of these scenarios.

\section{DDoS Detection and Localization}
\label{sec:methodology}

\begin{figure*}[htb]
\centering
\includegraphics[width=\textwidth]{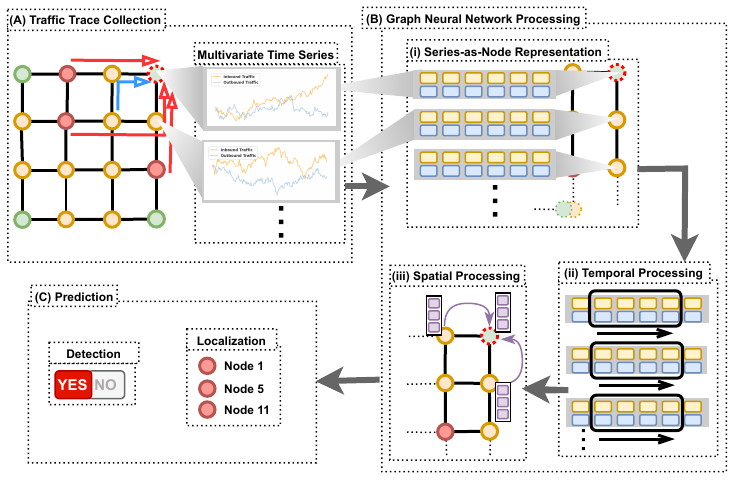}
\caption{Overview of the proposed DDoS detection and localization method at runtime: The traffic trace from each node, a spatiotemporal multivariate time series, is collected and transformed into a series-as-node in the graph. The graph is then processed by a GNN temporally and spatially (across nodes), which detects the attack and identifies the nodes with MIPs.}
\label{fig:overview}
\end{figure*}

Figure~\ref{fig:overview} provides an overview of the GNN-based DDoS detection and localization framework at \textit{runtime}. Inbound and outbound traffic traces from each router create a two-variable time series, which when combined, form a spatiotemporal multivariate time series across the NoC. This data structure preserves temporal patterns and spatial correlations (e.g., routing dependencies or cross-node congestion propagation), both of which are crucial for detecting distributed and stealthy attack behaviors. These traces are then represented as a graph and processed by a pre-trained GNN at \textit{runtime}. Each node in the graph represents a physical router or IP, and each feature vector captures the time series behavior for that node, enabling the GNN to treat DDoS detection as a dynamic graph classification task. The GNN makes two predictions: (1) whether a DDoS attack is occurring (detection), and (2) which MIPs are launching the attack (localization).

Note that the GNN is trained at \textit{design time} and then used for prediction at \textit{runtime}. During training, the GNN learns to correlate patterns in time series behavior with attack signatures The trained model is stored in the \textit{Security Engine (SE)}, an IP block with ML accelerator responsible for SoC security. The SE can be implemented as a lightweight neural processing unit (NPU) or embedded inference engine optimized for low-latency predictions. Upon localization, the SE enforces restrictions on MIPs network access to maintain system availability and notifies system administrators for further action.

\subsection{Traffic Trace Collection}
\label{sub-sec:traffic-trace-col}

To detect and localize DDoS attacks, we collect traffic traces from each router over a time interval, as shown in Figure~\ref{fig:overview}(A). The traces for each router consist of two inter-flit delay arrays, which represent the number of cycles between consecutive flits. These arrays are: inbound ($IIFD_j = \{i_0, i_1, \dots , i_{T_I}\}$) and outbound ($OIFD_j = \{o_0, o_1, \dots , o_{T_O}\}$), representing the inbound and outbound traffic, respectively. Together, they form a two-variable time series for each router.  Here, the subscript $j$ represents the router ID, $i_t$ represents the delay between $(t-1)^{th}$ and $t^{th}$ incoming flit, and $o_t$ represents the delay between $(t-1)^{th}$ and $t^{th}$ outgoing flit. $T_I$ and $T_O$ denote the total number of incoming and outgoing flits, respectively. At specified intervals, all inter-flit delays are gathered into an array and sent to the \textit{Security Engine}. 

Unlike~\cite{sudusinghe2021denial} and~\cite{sinha2021sniffer}, which rely on engineered features, our model uses raw features, such as inter-packet delays, for a more granular, real-time analysis of NoC traffic patterns. This allows the model to adapt to subtle, context-specific behaviors in traffic that are not being be captured by hand-crafted metrics. By using raw features, the GNN can automatically learn complex patterns, potentially uncovering anomalies that traditional ML models may overlook without extensive feature engineering. Additionally, this makes the approach inherently more flexible and transferable across different NoC architectures without the need for re-designing the feature extraction pipeline.

Two major hardware modifications are needed to support traffic collection and transmission to the \textit{SE}. First, we introduce a lightweight, event-triggered trace collection unit in each router. This unit includes a probe for incoming buffers, two 8-bit counters to track cycles between inbound flits and outbound flits, and packetizing logic. The packetizing logic compacts eight recent counter values into a single 1-flit packet, minimizing performance overhead. A similar counter-based mechanism has been used in hardware Trojans for NoC attacks~\cite{dhavlle2023defense}, where minimal overhead is key to avoiding detection. Our analysis shows less than 0.1\% area overhead for the modifications on a 64-node NoC-based SoC. 

We consider two methods for sending traffic traces from each router to the \textit{SE}: (1) a separate virtual network or (2) a separate physical network. While virtual networks require additional header fields and buffer space, which can impact MPSoC performance, separate physical NoCs add area overhead. However, advances in manufacturing have minimized wiring overhead. Given the limitations of on-chip buffer space and the complexity of managing virtual networks, we opt for the use of two physical NoCs. Yoon et al.~\cite{yoon2013virtual} demonstrated a 7\% area and power overhead for two physical NoCs, compared to 6\% for a single one, making the trade-off minimal. Using a separate physical NoC for trace collection ensures deterministic delivery of traffic metadata even under severe congestion, which is critical for maintaining observability during a DDoS attack. This approach also prevents interference in traffic trace transmission caused by ongoing DDoS attacks. The preceding discussion focuses on traffic trace collection at \textit{runtime}. However, the GNN is trained at \textit{design time}, requiring a large dataset for effective learning. To generate this dataset, we used a simulator configured to emulate the actual NoC-based SoC, with varying application mappings to produce diverse traffic patterns. Details of the simulator and data collection are discussed in Section~\ref{sec:experiments}.

\subsection{Graph Neural Network (GNN) Processing}
\label{subsec:gnn-proc}

GNNs are deep neural networks designed to handle graph-structured data, where nodes represent entities (e.g., NoC routers) and edges capture their relationships (e.g., NoC links or TSVs). GNNs have gained popularity for multivariate time series forecasting, imputation, anomaly detection, and classification~\cite{jin2024survey}. There are three key reasons for our selection of using GNNs for DDoS attack detection and localization in NoC. (1) GNNs, as deep learning models, eliminate the need for labor-intensive feature engineering, unlike in~\cite{sinha2021sniffer, sudusinghe2021denial}, and can capture complex relationships, leading to better generalization and robustness across diverse traffic patterns. (2) GNNs provide superior spatial awareness by inherently modeling the spatial relationships between nodes in the NoC as a graph and leveraging topology-aware predictions. (3) GNNs naturally handle both spatial and temporal dependencies in NoC traffic, making them ideal for accurately localizing malicious nodes over time. Furthermore, the inherent permutation invariance of GNNs enables robustness to the ordering of nodes, making the method adaptable to different layouts and core mappings in heterogeneous SoCs.    

\vspace{0.05in}
\noindent
\textbf{Graph Representation of Traffic Traces:}
A time series is a sequence of data points that occur in successive order over some period of time. The GNN receives a two-variable ($IIFD_j$, $OIFD_j$) time series from each router. Together, the data from all routers forms a spatiotemporal multivariate time series. Here, ‘spatial’ refers to traffic data collected across all routers in the NoC, and ‘multivariate’ indicates that the time series captures multiple variables (for inbound and outbound traffic). Let the spatiotemporal multivariate time series be represented by the matrix $\mathbf{Y} \in \mathbb{R}^{2N \times T}$, where $N$ is the number of routers (nodes) in the NoC, and $T$ represents the total flits captured in the inter-flit delays for each direction at a single node. For each router $j \in \{1, 2, \dots, N\}$, the matrix $\mathbf{Y}$ includes two variables: $IIFD_j$ for inbound traffic and $OIFD_j$ for outbound traffic. This means the complete traffic trace of the NoC represents a multivariate time series with $2N$ variables collected over a total of $T$ flits. This structured representation enables consistent graph construction independent of router count or physical layout, which facilitates scalable deployment.

A spatiotemporal graph with fixed structure for a traffic trace ($\mathbf{Y}$) can be defined as $\mathcal{G} = \{ \mathcal{G}_1, \mathcal{G}_2, \dots, \mathcal{G}_T \}$, where $\mathcal{G}_t = (\mathbf{A}, \mathbf{X_t})$ denotes an attributed graph at time $t$. $\mathbf{A} \in \mathbb{R}^{N \times N}$ and $\mathbf{X_t} \in \mathbb{R}^{N \times 2}$ are the corresponding adjacency and feature matrices. The adjacency matrix ($\mathbf{A}$) represents the graph topology and is fixed over time. We use pair-wise connectivity to construct the adjacency matrix as follows:
\[
\mathbf{A}_{i,j} = 
\begin{cases} 
1, & \text{if } r_i \text{ and } r_j \text{ are directly linked}, \\
0, & \text{otherwise}.
\end{cases}
\]

Here, $r_i$ and $r_j$ correspond to routers in the NoC. Our graph representation of the traffic traces follows the same topology as the routers in the NoC ($n \times n$ mesh), enabling topology-aware prediction. The feature matrix $\mathbf{X_t}$ represents the traffic data at each time step $t$, where each node contains two variables: the inbound and outbound inter-flit delays. Thus, this representation follows a series-as-node model, where each of the $N$ nodes in the graph corresponds to a router. For each router $j$, the two-variable time series consists of the inbound ($IIFD_j$) and outbound ($OIFD_j$) inter-flit delays, and this time series is assigned to the corresponding node in the graph. A section of this graph is visualized in Figure~\ref{fig:overview}(B)(i). This design ensures that traffic dynamics are encoded at the granularity of each router, offering fine resolution for node-wise classification.

\vspace{0.05in}
\noindent
\textbf{GNN Architecture:} We use a \textit{spatial GNN} to process spatial-temporal graphs ($\mathcal{G}$s) to detect and localize DDoS attacks. Specifically, a \textit{spatial GNN} focuses on processing the graph by passing messages between connected nodes to learn representations for each node based on its neighbors. The GNN processes a traffic trace graph ($\mathcal{G}$) in two ways: \textit{temporally} over IFDs and \textit{spatially} across routers (nodes in the NoC). \textit{Temporal processing} is achieved by analyzing the graph over consecutive inter-flit delays, as illustrated in Figure~\ref{fig:overview}(B)(ii). At each time step $t$, the graph $\mathcal{G}_t$ is updated with new traffic data, capturing the evolving patterns over time. \textit{Spatial processing} involves aggregating information from neighboring nodes as visualized in Figure~\ref{fig:overview}(B)(iii). Together, these mechanisms allow the model to detect subtle deviations in local traffic that correlate with global congestion or attack patterns.

For a node $v_i$, the message passing process can be described as:
\[
\mathbf{a}_i^{(k)} = \text{AGGREGATE}^{(k)} \left( \{ \mathbf{h}_j^{(k-1)} \mid j \in \mathcal{N}(i) \} \right)
\]

Here, $\mathbf{h}_j^{(k-1)}$ represents the feature embedding of neighboring node $v_j$ from the previous GNN layer $k-1$, containing the state of node $j$ before the current message-passing round. The \textit{AGGREGATE} function collects and aggregates these messages from neighboring nodes to generate $\mathbf{a}_i^{(k)}$, the aggregated message for node $v_i$. In the \textit{COMBINE} step, the aggregated message $\mathbf{a}_i^{(k)}$ is combined with the previous state of node $v_i$, represented by $\mathbf{h}_i^{(k-1)}$, to update its feature embedding for the current layer $k$:
\[
\mathbf{h}_i^{(k)} = \text{COMBINE}^{(k)} \left( \mathbf{h}_i^{(k-1)}, \mathbf{a}_i^{(k)} \right)
\]

Here, $\mathbf{h}_i^{(k-1)}$ represents the feature embedding of node $v_i$ from the previous layer, while $\mathbf{a}_i^{(k)}$ is the newly aggregated message from its neighbors. The \textit{COMBINE} function integrates these two to produce the updated node embedding $\mathbf{h}_i^{(k)}$. These embedding updates progressively refine the node representation, enabling discrimination between benign and malicious behavior even when traffic patterns overlap.

The GNN architecture consists of $n_{\text{conv}}$ 1D convolution layers for temporal processing, followed by $n_{\text{gr}}$ layers of either Graph Convolutional Network (GCNConv), Graph Attention Network (GATConv), or GraphConv~\cite{morris2019weisfeiler} for spatial processing (message passing and feature aggregation from neighboring nodes), and finally $n_{\text{fc}}$ fully connected layers for classification. For temporal processing layer, several options were considered, including recurrent layers, convolutional layer, hybrid mix of above two, and transformer architecture. While recurrent layers excel at learning temporal patterns, they suffer from scalability issues and have longer training and inference times. Transformers, though powerful, introduce significant overhead, making them impractical for a lightweight model. Convolutional layers are faster on accelerators for both training and inference, making them the preferred choice for quick and accurate temporal processing. In our model, each convolutional layer is followed by a pooling layer to downsample the input, reduce dimensions, and control overfitting. Dropout layers are utilized to prevent overfitting. The input to the model is a two-variable, $l$-length time series ($IIFD_j$ and $OIFD_j$) for each node, extracted from $\mathcal{G}$. Hyperparameter tuning was used to finalize the number of layers and parameters, as detailed in Section~\ref{sec:experiments}. This architectural design ensures that the model remains compact, interpretable, and efficient for integration in SoC security IPs.

\subsection{GNN Training and Prediction}

The GNN model is tackling two key problems: DDoS detection and localization. Given a spatiotemporal graph ($\mathcal{G}$) as traffic trace input, the GNN predicts whether a DDoS attack is occurring (detection) and which nodes are malicious (localization). Attack detection is formulated as a graph classification problem, whereas localization is treated as a node classification problem. This dual-task formulation allows the model to learn fine-grained per-node behavior while also capturing high-level attack presence across the network. 

The GNN is trained by minimizing the localization loss. Binary cross-entropy is employed for training the localization task, as it is a binary classification problem on each node. While the model is explicitly supervised only for localization, detection is implicitly derived based on the node-level outputs, as discussed later. The model uses the Adam optimizer (initial learning rate: 0.0005) with a scheduled 0.1 rate reduction after 15 epochs if validation loss plateaus. Early stopping was used to halt training if validation loss does not improve for 60 epochs. Batch size and other model parameters were tuned via hyperparameter optimization, as detailed in Section~\ref{subsec:hyperparm}. Training was conducted over several randomized seeds and traffic scenarios to ensure generalization across MIP placement, topology variance, and application mappings.

Algorithm~\ref{alg:det_and_loc} summarizes the inference for detection and localization. The pre-trained GNN receives input features as a matrix $\mathbf{Y}$, multivariate time-series representing traffic. In line 2, the matrix $\mathbf{Y}$ is converted into a graph $\mathcal{G}$, as discussed in Section~\ref{subsec:gnn-proc}. Line 3 shows node classification returning a vector ($\mathbf{n\_pred}$) of length $N$, where each element is $1$ (malicious node) or $0$ (benign node). Regardless of being trained only on localization, our approach performs inference for both detection and localization. Lines 4-7 compute the graph prediction: the graph is classified as an attack (line 4 \& 5) if at least one MIP is localized (i.e., if $\mathbf{n\_pred}$ contains at least one ``1"). Otherwise, it is classified as normal (line 7). Finally, $\mathbf{g\_pred}$ and $\mathbf{n\_pred}$ are returned, providing predictions for detection and localization, respectively. 

This inference method offers a lightweight post-processing logic that does not require additional parameters or a separate classifier for detection. It also reflects real-world defensive action such that identification of even one compromised node is sufficient to trigger mitigation mechanisms.

\begin{algorithm}
\caption{Inference for Detection and Localization }
\label{alg:det_and_loc} 
\begin{algorithmic}[1]
\Function{\textit{detectAndLocalize}}{$\textbf{Y}$}
\State $\mathcal{G}$ $\leftarrow$ \texttt{representAsGraph($\textbf{Y}$)}
\State $\mathbf{n\_pred} \leftarrow \texttt{GNNInference($\mathcal{G}$)} \; \text{where} \; \mathbf{n\_pred} = [b_1, b_2, \dots, b_N], \; b_i \in \{0, 1\}$
\If{$\sum_{i=1}^N b_i \geq 1$}
    \State $\mathbf{g\_pred} \leftarrow 1$ \Comment{An attack detected}
\Else
    \State $\mathbf{g\_pred} \leftarrow 0$ \Comment{No attack detected}
\EndIf
\State \textbf{retrun} $\mathbf{g\_pred}, \mathbf{n\_pred}$
\EndFunction
\end{algorithmic}
\end{algorithm}

Both detection and localization serve as critical pillars in a complete NoC security framework. Detection provides a global signal for system-wide threat assessment, while localization enables precise, node-level countermeasures to isolate or restrict attacker behavior. We evaluate both detection and localization performance using accuracy, precision, recall, and F1-score. 

For detection, accuracy is the ratio of correctly classified graphs (attack or normal) to the total number of graphs. In this context, a true positive (TP) refers to an attack graph correctly identified as an attack, while a true negative (TN) denotes a normal graph correctly identified as normal. For localization, accuracy is the proportion of correctly identified nodes (malicious or benign) to the total number of nodes. Here, a true positive is a malicious node correctly identified as malicious, and a true negative is a benign node correctly identified as benign. We also report precision and recall to capture the trade-off between false alarms and missed detections, particularly under class imbalance where benign nodes significantly outnumber malicious ones.

\section{Experiments}
\label{sec:experiments}

In this section, we first describe the experimental setup, followed by data collection and hyperparameter tuning. Next, we present the results for detection and localization. We evaluate the proposed approach using multiple synthetic and benchmark traffic patterns in both 2D and 3D mesh-based NoC topologies. The experiments are designed to assess performance across different attack intensities, router placements, and interconnect structures. Finally, we evaluate the robustness of our approach. Our evaluation criteria include detection accuracy, localization precision, scalability, and response consistency across multiple scenarios.

\begin{table}[tbp]
% \vspace{-0.1in}
\caption{System and interconnect configurations.}
% \vspace{-0.15in}
\label{tab:configTable}
\begin{center}
\begin{tabular}{|p{0.50\columnwidth} | p{0.40\columnwidth}|}
\hline
\textbf{Parameter} & \textbf{Details} \\
\hline
Number of cores & 64  \\
\hline
2D Topology & 8x8  \\
\hline
3D Topology & 4x4x4 \\
\hline
Processor architecture & X86,\\
\hline
Processor frequency & 2GHz \\
\hline
Cache coherency protocol & MESI two-level\\
\hline
L1 instruction \& data cache & 32KB, 32KB (private) \\
\hline
L2 cache & 512KB (shared) \\
\hline
Routing protocol & XY deterministic\\
\hline
\end{tabular}
\label{tab1}
\end{center}
% \vspace{-0.1in}
\end{table}

\subsection{Experimental Setup}

To evaluate the effectiveness of our DDoS detection and localization approach, we simulated a 64-node MPSoC using both the cycle-accurate gem5~\cite{gem5} and Noxim~\cite{catania2016cycle} simulators. Full system simulations were performed in gem5 to collect network traces, utilizing seven benchmarks from SPLASH-2~\cite{woo1995splash} and PARSEC~\cite{bienia2008parsec}: \textit{fft}, \textit{fmm}, \textit{lu}, \textit{barnes}, \textit{radix}, \textit{blackscholes}, and \textit{ocean}. These benchmarks were selected due to their varied memory access patterns and inter-core communication intensities, which allowed us to simulate diverse traffic behaviors in the NoC. Noxim was modified to enable both trace-based and table-based simulations to run simultaneously. Trace-based traffic, generated from gem5, was used to represent legitimate applications running on the MPSoC, while table-based traffic simulated flooding by MIPs, targeting selected nodes with memory controllers, to mimic a DDoS attack. Flooding traffic was synthesized with varying packet injection rates and randomized source nodes to mimic coordinated attacks under different intensities and spatial distributions. 

For 3D NoC-based experiments, we modified Noxim to support TSVs and use the same rest of the setup to conduct 3D NoC experiments. The vertical inter-layer links were modeled using a shared TSV arbitration mechanism, where multiple layers compete for a single vertical path at each (x, y) location. The 2D experiments were done in 8x8 topology with four memory controllers at each corner while 3D were done in a 4x4x4 topology with four memory controllers at alternating corners of top and bottom layers. For consistency, benchmark mapping and attack scenarios were matched across 2D and 3D setups to ensure fair comparison. Details of the configurations used in the simulations are provided in Table~\ref{tab:configTable}.

\subsection{Training Data Collection}
\label{subsec:data-col}

We consider two scenarios for training data collection: attack and normal. In the attack scenario, one or more MIPs are launching DDoS attacks, while the normal scenario involves no such attacks. We collected multiple traffic traces ($\mathbf{Y}$s) by changing the mapping of the application running on the system and randomly selecting MIP nodes. For each mapping, we collect six independent traffic traces: two normal and four attack scenarios. This approach ensures balanced coverage across application behavior and attacker configurations.

The two normal traces are obtained by monitoring the system over an extended period and dividing the data into two segments, each of length $l$. In three of the attack scenarios, $N_M$ randomly selected MIPs target one VIP, while in the fourth scenario, $N_M$ MIPs target two VIPs. These configurations reflect common DDoS structures, coordinated attack focus and distributed attack dispersion. For the default experiment, we use 3 MIPs ($N_M=3$), each with a packet injection rate (PIR) of 0.05. A PIR of 0.05 was chosen based on our observation that three MIPs can trigger an attack condition ($<$ 30\% increase in starved flits) at a PIR of 0.04 or higher. This threshold aligns with typical starvation-induced bottlenecks observed in realistic NoC traffic scenarios.

We evaluate the model's performance by varying these parameters in Section~\ref{subsec:robustness}. The time series length $l$ was set to 400, selected from 200, 400, and 600, based on training curve observations of GNN performance. We observed that shorter sequences limit the learning of longer-term dependencies, while longer sequences lead to higher variance and increased model complexity.

The full dataset consists of 2688 traces ($Y$s) across 7 benchmarks, with each trace corresponding to a graph representation ($\mathcal{G}$). We split the dataset into 90\% for training and 10\% for testing. The imbalance between attack and normal scenarios (1:2) does not directly affect the detection task (graph classification) since the GNN is trained on localization. However, for the localization task (node classification), there is an inherent imbalance between the two classes (malicious vs. benign nodes), because the number of MIPs is always smaller than the total number of IPs ($N_M < N$). To mitigate this issue, class weights are calculated based on the frequency of class labels, and these weights are applied during loss calculation via binary cross-entropy loss function. This weighting ensures that the model remains sensitive to the underrepresented malicious class without overfitting, improving both precision and recall.

\subsection{Hyperparameter Tuning}
\label{subsec:hyperparm}

We conducted rigorous hyperparameter tuning using both grid search and random search. Table~\ref{tab:hyperparam} outlines each hyperparameter, its search space, and the selected values. The final model consists of 10 layers: four convolution layers, two spatial processing layers, and four fully connected layers. This architecture was selected to strike a balance between representation capacity and training stability. Each component was optimized to address either spatial or temporal complexity in the input data.

After testing with Graph Convolutional Networks (GCN), Graph Attention Networks (GAT), and GraphConv~\cite{morris2019weisfeiler}, GraphConv was selected for its superior efficiency and accuracy in localization. Specifically, GraphConv achieved 2--4\% higher localization F1-score on average compared to GCN and GAT, while also requiring fewer parameters and offering lower inference latency, which is an important consideration for real-time operation in constrained Security Engines. Its simplicity also reduces the need for tuning attention weights or layer normalization parameters, making it more deployment-friendly. GAT offered competitive accuracy but had higher training complexity due to attention mechanisms, making GraphConv a more balanced choice. Similarly, we explored different temporal processing strategies, including recurrent layers (LSTM), Transformer-based encoders, and 1D convolutional layers. While Transformer models slightly outperformed convolution layers in accuracy under ideal conditions, they incurred significantly longer training times and higher memory requirements. Moreover, they required larger datasets to generalize well, which can be impractical in runtime-adaptive security engines. Recurrent layers struggled with long sequences due to vanishing gradients. 1D convolutional layers offered the best balance between performance and computational efficiency, especially for training and inference on accelerators. Thus, they were selected for temporal modeling. Convolutions also allow for parallel processing of time steps, which significantly reduces inference latency—critical in attack response scenarios.

Similarly, average pooling was selected against max pooling. We found that average pooling produced more stable performance, especially under bursty traffic patterns, where max pooling tended to amplify noise. This was particularly useful when modeling traces under rapid shifts in PIR, where spike patterns may not be consistent across nodes. The selected kernel sizes and strides vary by layer. For example, the convolution kernel sizes ([5, 10, 10, 10]) indicate the first convolution layer uses a kernel size of 5, while the others use a kernel size of 10. Convolution strides, and pooling kernel sizes and strides follow the same layer order. Dropout is applied after each convolution layer (30\%) and after the second FC layer (50\%) to prevent overfitting. The four fully connected layers have 400, 133, 44, and 1 neuron, respectively. Increasing the number of spatial layers beyond two led to minor improvements in accuracy but also caused overfitting on the training data. This was validated through ablation studies where deeper GNNs exhibited higher training accuracy but lower generalization on unseen mappings. The chosen configuration reflects the best trade-off between generalization, accuracy, and runtime feasibility, as confirmed by our ablation results.

\begin{table}[tbp]
\centering
\caption{Hyperparameter tuning and selected values for GNN model.}
\vspace{-0.05in}
\label{tab:hyperparam}
\resizebox{0.9\columnwidth}{!}{%
\begin{tabular}{|l|l|l|}
\hline
\textbf{Hyperparameter} & \textbf{Search Space} & \textbf{Selected Value} \\ \hline
\# conv layers ($n_{conv}$) & 2, 3, 4, 5 & 4 \\ \hline
\# spatial proc. layers ($n_{gr}$) & 1, 2 & 2 \\ \hline
\# FC layers ($n_{fc})$ & 3, 4, 5 & 4 \\ \hline
Spatial proc. layer type & GCN, GAT, GraphConv & GraphConv \\ \hline
Type of pooling & Average, Max & Average\\ \hline
Conv kernel size & 5, 10, 15, 20, 25 & [5, 10, 10, 10] \\ \hline
Conv stride & 1, 2 & [1, 1, 1, 2] \\ \hline
Pool kernel size & 5, 10, 15 & [5, 5, 5, 5] \\ \hline
Pool stride & 1, 2 & [1, 2, 2, 2]\\ \hline
Batch size & 2,4,8,16,32,64,72,80,88,128 & 64 \\ \hline
FC dropout rate & 40\%, 50\%, 60\%, 70\% &  50\% \\ \hline
Conv dropout rate & 10\%, 20\%, 30\%, 40\% &  30\% \\ \hline
% No. of FC dropout layers & 1, 2, 3, 4 &  1 \\ \hline
% No. of Conv dropout layers & 1, 2, 3, 4 &  4 \\ \hline

\end{tabular}%
}
% \vspace{-0.2in}
\end{table}

\begin{table}[b]
\centering
\caption{DDoS detection and localization with 3 MIPs.}
\vspace{-0.05in}
\label{tab:results-main}
\begin{tabular}{l|l|l|l|l|}
\cline{2-5}
\textbf{} & \textbf{Accuracy} & \textbf{Precision} & \textbf{Recall} & \textbf{F1-score} \\ \hline
\multicolumn{1}{|l|}{\textbf{Detection}} & 100.00\% & 100.00\% & 100.00\% & 100.00\% \\ \hline
\multicolumn{1}{|l|}{\textbf{Localization}} & 99.07\% & 89.58\% & 91.11\% & 90.34\% \\ \hline
\end{tabular}%
\end{table}

\subsection{DDoS Detection and Localization Performance}

Table~\ref{tab:results-main} presents the accuracy, precision, recall, and F1 score for both detection and localization of DDoS attacks. These experiments were conducted using the default parameters ($N_M=3$, $PIR=0.05$) outlined in Section~\ref{subsec:data-col}. The detection task achieved perfect scores across all metrics, which does not indicate overfitting, as the model was trained to minimize localization loss rather than detection loss. The localization task also showed strong performance, with a high accuracy (99\%), demonstrating the model's robustness across varying traffic patterns and node behaviors. Even in the presence of background noise and minor traffic perturbations caused by concurrent applications, the GNN maintained consistent performance, showing resilience to non-malicious variation.

The slightly lower precision, recall, and F1-score compared to the exceptionally high accuracy can be attributed to class imbalance in a graph (more benign than malicious nodes), a common challenge in NoC security contexts. This is particularly significant in large networks where malicious nodes typically form a small fraction of the total node population. However, the recall remains sufficiently high, which is crucial for DDoS attack localization, ensuring that most MIPs are detected and making the approach reliable in critical scenarios involving real-time system protection. This high recall supports timely enforcement actions, reducing the window of vulnerability.

Table~\ref{tab:sota} summarizes a comparison with related works on DoS/DDoS attack detection and localization. Unlike some methods~\cite{sudusinghe2021denial, js2015runtime}, which primarily focus on single MIP attack scenarios or do not provide localization, our approach handles multiple MIP and VIP attacks in an 8x8 mesh using a topology-aware graph neural network. This enables our model to represent a broader range of attack surfaces, including coordinated and overlapping-path attacks that strain shared NoC resources. Our method achieves higher detection accuracy, outperforming previous works that either do not report accuracy or fall short in detection. 

Moreover, our approach also excels in localization, achieving higher accuracy than others, which either lack localization capabilities or report lower localization accuracy. Notably, the model's ability to precisely identify MIPs, even in scenarios with multiple attackers or distributed VIPs, allows for more granular and effective defense mechanisms. In addition to supporting complex attack scenarios, our framework is the \textit{only one} among the compared methods that generalizes to both 2D and 3D NoC architectures, including evaluations on a 4$\times$4$\times$4 3D mesh topology with Through-Silicon Vias (TSVs). This 3D adaptability ensures applicability to emerging stacked SoC designs with vertical inter-layer communication, which is increasingly common in heterogeneous architectures. Furthermore, while prior approaches such as~\cite{sinha2021sniffer, wang2024dl2fence} either rely on local heuristics or domain-specific pre-processing (e.g., CNN with frame fusion), our solution operates directly on raw traffic traces using spatiotemporal multivariate time series, eliminating the need for manual feature engineering or handcrafted segmentation. This reduction in pre-processing complexity also lowers integration effort and makes the solution more agile in dynamic systems. Our methodology also unifies detection and localization through a single GNN pipeline, whereas other works either omit one of the tasks or decouple them into separate modules, increasing overhead and limiting real-time adaptability. By framing detection and localization as graph-level and node-level classification problems respectively, we achieve a clean abstraction that promotes both accuracy and deployment simplicity. Overall, our approach achieves state-of-the-art performance, is generalizable across topologies and attack configurations, and introduces minimal hardware overhead, making it a practical and scalable security solution for modern and future NoC-based SoCs.

\begin{table*}[htp]
\caption{Comparison of related works on DoS/DDoS detection and localization in NoCs. Each row represents a method and each column is a key evaluation criterion.}
% \vspace{-0.05in}
\label{tab:sota_transposed}
\label{tab:sota}
\resizebox{\textwidth}{!}{%
\begin{tabular}{l|c|c|c|c|c|c}
\hline
\textbf{Work} &
\textbf{Attack Scenarios} &
\textbf{Mesh Size} &
\textbf{Detection Method} &
\textbf{Detection Accuracy} &
\textbf{\begin{tabular}[c]{@{}l@{}} Localization \\  Method \end{tabular}} &
\textbf{\begin{tabular}[c]{@{}l@{}} Localization  \\  Accuracy \end{tabular}} \\
\hline
\cite{js2015runtime} &
single MIP &
8x8 &
\begin{tabular}[c]{@{}l@{}} runtime latency \\  auditor \end{tabular}  &
not provided &
no localization &
no localization \\
\hline
\cite{charles2020real} &\begin{tabular}[c]{@{}l@{}} multiple MIP \& \\  multiple VIP \end{tabular}  &
4x4, 8x8 &
traffic monitoring &
not provided & \begin{tabular}[c]{@{}l@{}} traffic monitoring \& \\  message passing  \end{tabular} &
not provided \\
\hline
\cite{sudusinghe2021denial} &
single MIP &
4x4 &
via machine learning &
94.95\% -- 98.93\% &
no localization &
no localization \\
\hline
\cite{sinha2021sniffer} &
\begin{tabular}[c]{@{}l@{}} multiple MIP \& \\  multiple VIP \end{tabular} &
8x8 & 
\begin{tabular}[c]{@{}l@{}} perceptrons for local \\  congestion detection\end{tabular}  &
97.52\% &
\begin{tabular}[c]{@{}l@{}} congestion sharing \& \\  collective decisioning \end{tabular} &
96.75\% \\
\hline 
~\cite{wang2024dl2fence} &
\begin{tabular}[c]{@{}l@{}} multiple MIP \& \\  multiple VIP \end{tabular} &
8x8 &
CNN with frame fusion &
$\sim$99\%  &
\begin{tabular}[c]{@{}l@{}} segmentation via \\  CNN decoder  \end{tabular}   &
$\sim$98.5\%  \\
\hline
\textbf{\begin{tabular}[c]{@{}l@{}}Our \\ Approach\end{tabular}} &
\begin{tabular}[c]{@{}l@{}} multiple MIP \& \\  multiple VIP \end{tabular} &
8x8, 4x4x4 &
\begin{tabular}[c]{@{}l@{}}via topology-aware \\ graph neural network\end{tabular} &
100\% &
\begin{tabular}[c]{@{}l@{}}via topology-aware \\ graph neural network\end{tabular}   &
99\% \\
\hline
\end{tabular}%
}
% \vspace{-0.1in}
\end{table*}

\subsection{Robustness of DDoS Detection and Localization }
\label{subsec:robustness}

To evaluate the robustness and generalization ability of our GNN-based detection and localization framework, we perform three sets of experiments: (i) increasing the number of malicious IPs (MIPs), (ii) varying the packet injection rate (PIR), and (iii) generalizing to a 3D NoC topology with through-silicon vias (TSVs). These experiments assess how the model adapts to dynamic adversarial strategies and evolving system architectures, which are common in real-world SoC deployments.

\subsubsection{Impact of the Number of Malicious IPs}
\label{subsubsec:mip_impact}

\begin{figure}[htp]
\centering
\includegraphics[width=\columnwidth]{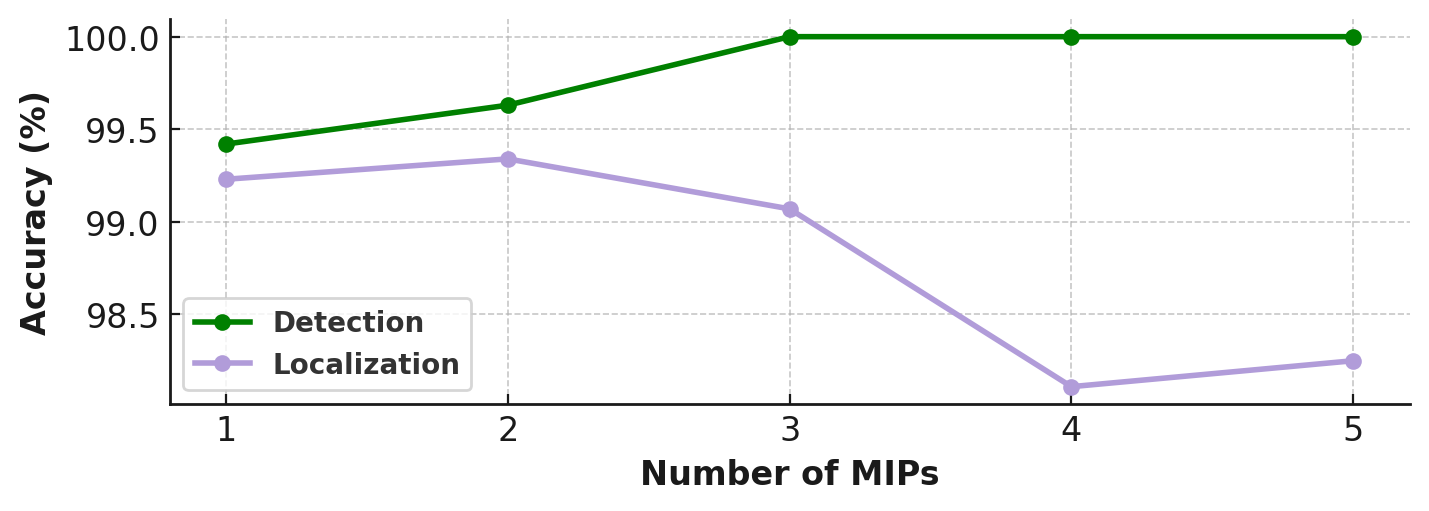}
\caption{Accuracy of detection and localization for increasing number of malicious IPs (MIPs).}
\label{fig:mip_graph}
\end{figure}

Figure~\ref{fig:mip_graph} shows the detection and localization accuracy as the number of malicious IPs increases from 1 to 5. We observe that the detection accuracy remains consistently high (above 99.5\%) and saturates at 100\% beyond 3 MIPs. This shows that the model is highly effective in identifying abnormal traffic patterns introduced by multiple distributed sources. The stable detection curve demonstrates that the model does not rely on specific MIP placements or synchronized injection patterns. 

However, localization accuracy slightly drops as the number of MIPs increases. This is likely due to the distributed nature of coordinated DDoS attacks: when more nodes are involved, the malicious traffic becomes more spread out and diluted, reducing the distinguishable impact of individual MIPs. For instance, at 4 and 5 MIPs, the localization accuracy dips to 98.2\%, though it still remains highly reliable. This trade-off is expected as attack footprint dispersion makes isolating exact contributors more challenging. Nonetheless, the degradation is minimal and well within acceptable limits for practical defense mechanisms.

\subsubsection{Impact of Packet Injection Rate}
\label{subsubsec:pir_impact}

\begin{figure}[htp]
\centering
\includegraphics[width=\columnwidth]{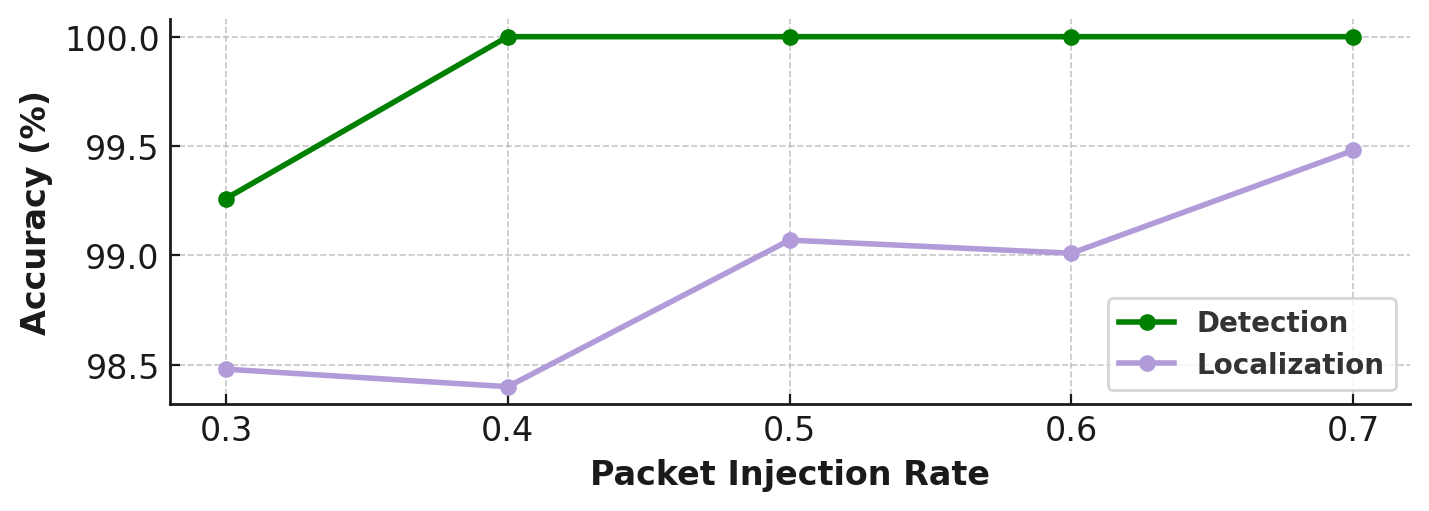}
\vspace{-0.2in}
\caption{Accuracy of detection and localization for increasing packet injection rate (PIR) by MIPs.}
\label{fig:pir_graph}
\end{figure}

We next examine how the model behaves under different attack intensities, controlled via the packet injection rate (PIR) of 3 malicious nodes. Figure~\ref{fig:pir_graph} illustrates the detection and localization performance for PIR values ranging from 0.3 to 0.7. Detection remains perfect (100\%) for PIR $\geq$ 0.4, while at PIR = 0.3, it slightly dips to 99.3\%. This reflects the model's sensitivity to traffic volume and the visibility of attack-induced congestion in the NoC.

Localization shows a more noticeable variance: at lower PIRs, fewer bogus packets are injected, leading to weaker disruption signals and hence lower classification confidence. At higher PIRs, the disruption becomes more evident, increasing localization precision as the model identifies strong deviation signatures across time and nodes. This result confirms that the framework is sensitive to subtle patterns even at low PIRs, yet naturally benefits from higher traffic volumes.

\subsubsection{3D NoC with TSVs}
\label{subsubsec:3dnoc_impact}

\begin{table}[b]
\centering
\caption{DDoS detection and localization with 3 MIPs in 3D NoC.}
\vspace{-0.05in}
\label{tab:results-main-3D}
\begin{tabular}{l|l|l|l|l|}
\cline{2-5}
\textbf{} & \textbf{Accuracy} & \textbf{Precision} & \textbf{Recall} & \textbf{F1-score} \\ \hline
\multicolumn{1}{|l|}{\textbf{Detection}} & 100.00\% & 100.00\% & 100.00\% & 100.00\% \\ \hline
\multicolumn{1}{|l|}{\textbf{Localization}} & 98.12\% & 90.15\% & 90.31\% & 89.37\% \\ \hline
\end{tabular}%
\end{table}

To evaluate generalization to more complex architectures, we extend our evaluation to a $4 \times 4 \times 4$ 3D mesh NoC with vertical links (TSVs). The data collection and processing methodology remain consistent with the 2D NoC setup, with two key changes: (1) the underlying topology is a 3D graph constructed by connecting vertical neighbors using TSVs, and (2) the graph structure includes additional Z-axis links during adjacency matrix generation. These changes reflect architectural enhancements increasingly adopted in high-performance stacked SoCs. 

The rest of the GNN processing pipeline remains unchanged. For the GNN structure, we conduct use the same selected values for hyperparameters as mentioned in Table~\ref{tab:hyperparam}. Table~\ref{tab:results-main-3D} summarizes the performance of the proposed model under this new topology. Detection remains flawless across all metrics (accuracy, precision, recall, F1-score = 100\%). Localization also performs well, with 98.12\% accuracy and over 90\% on all classification metrics. This indicates that the model’s topology-aware design extends naturally to higher-dimensional networks, without requiring architectural re-training or topology-specific feature engineering. These results confirm that the proposed GNN framework generalizes well to 3D NoCs, validating its scalability and robustness to structural complexity. These three experiments collectively demonstrate the robustness of our proposed model under varying adversarial conditions and architectural complexities. The performance remains consistently high even under increased attack complexity, lower traffic injection, and extended topologies such as 3D NoCs. This positions our GNN-based approach as a highly scalable and resilient solution for future heterogeneous SoC designs.

\section{conclusion}
\label{sec:conclusion}

Securing on-chip communication is crucial for trustworthy electronic systems, particularly in the presence of sophisticated and distributed threats. While existing techniques for mitigating Denial-of-Service (DoS) attacks in Network-on-Chip architectures provide some level of protection, they often fall short when faced with Distributed DoS (DDoS) attacks that originate from multiple malicious sources and exhibit dynamic, stealthy traffic patterns. In this paper, we presented a topology-aware framework that employs Graph Neural Networks (GNNs) to detect and localize DDoS attacks in NoCs by modeling the chip's communication fabric as a graph. This approach enables effective analysis of spatiotemporal traffic behaviors, leveraging both structural connectivity and temporal variations to accurately classify and pinpoint malicious activity. Our extensive experimental evaluation shows that the proposed method achieves up to 100\% detection accuracy and 99\% localization accuracy. Unlike previous methods that rely on static heuristics or engineered features, our GNN-based solution adapts directly to raw inter-flit delay data, providing better resilience and insight into fine-grained communication anomalies. Specifically, the model remains robust under varying packet injection rates, increasing numbers of malicious IPs, diverse placements of attackers, and different application benchmarks. 

Moreover, we demonstrate that this framework scales beyond traditional 2D mesh NoCs and performs reliably in more complex 3D NoC architectures that incorporate Through-Silicon Vias, without the need for architecture-specific tuning or feature engineering. This highlights the flexibility of our method to operate in next-generation SoC environments that feature stacked memory and vertically integrated compute resources. The approach also eliminates reliance on manual thresholds or traffic templates, making it suitable for modern heterogeneous SoCs with dynamic runtime behavior. In addition to its detection capabilities, the system can be deployed with a lightweight security engine that introduces negligible hardware overhead, enabling practical integration into real-time runtime environments. By unifying detection and localization through a single model, we also reduce system complexity, improve response time, and minimize design-time efforts. Our results confirm that topology-aware GNN-based DDoS detection and localization is not only accurate but also scalable, generalizable, and hardware-friendly. This work lays the foundation for future research in applying graph learning methods to other runtime NoC threats such as covert channels, eavesdropping, or traffic shaping attacks in complex SoC designs.

% \textbf{Have the authors considered the impact of the noises in the event-triggered trace collection unit?}

% Yes, noise is essential for ensuring a realistic and comprehensive study. In our method, we account for noise in two ways: in collection and training. First, legitimate communication between nodes occurs simultaneously with the attacker’s Distributed Denial of Service (DDoS) attack, creating noise that complicates detection and localization. For instance, as shown in Figure 1, when Node 5 and 10 launches a DDoS attack, legitimate communication from Node 1 to Memory Controller 2 adds interference by sharing paths, buffers, and memory controllers with the malicious traffic. Additionally, we incorporate Gaussian noise during the training of our Graph Neural Network (GNN) to augment the data and prevent overfitting. This approach ensures the model remains robust and effective in handling noisy scenarios, both in training and real-world applications.

\color{black}

\bibliographystyle{IEEEtran}
\bibliography{IEEEabrv,bibliography.bib}

\end{document}